\documentclass{elsart}
\usepackage{latexsym}
\usepackage[final]{graphicx}
                                                                                
\usepackage{natbib}
\usepackage{amssymb}

\begin{document}

\begin{frontmatter}
\title{Distribution of Positron Annihilation Radiation\protect\footnote{Thanks
to J.D. Kurfess, M.D. Leising and G. Weidenspointner
for helpful discussions, and to the
NASA Grants DPR S-13801-G and NNG05GK37G for support.}
}

\author{Peter A. Milne}
\address{Department of Astronomy, University of Arizona, Tucson, AZ 85721}

\begin{abstract}
The SPI instrument on-board the ESA/INTEGRAL satellite is
engaged in a mission-long study of positron annihilation
radiation from the Galaxy. Early results suggest that the
disk component is only weakly detected at 511 keV by SPI.
We review CGRO/OSSE, TGRS and SMM studies of 511 keV line 
and positronium continuum emission from the Galaxy in light of the early 
INTEGRAL/SPI findings. We find that when similar spatial distributions 
are compared, combined fits to the OSSE/SMM/TGRS data-sets produce 
bulge and disk fluxes similar in total flux and in B/D ratio to the 
fits reported for SPI observations. We further find that the 511 keV 
line width reported by SPI is similar to the values reported by TGRS, 
particularly when spectral fits include both narrow-line and broad-line 
components. Collectively, the consistency between these four instruments 
suggests that all may be providing an accurate view of positron annihilation 
in the Galaxy.
\end{abstract}
\begin{keyword}
\PACS
\end{keyword}
\end{frontmatter}

\section{Introduction}

The successful launch and operation of the SPI instrument on-board 
ESA's INTEGRAL satellite has renewed interest in the study of positron 
annihilation in the Galaxy. Numerous publications have resulted from 
the analysis of subsets of the first years of the SPI data-set. Both 
the spatial distribution and the spectral features have been studied, 
and preliminary findings of the analysis of the positronium continuum 
component is currently in preparation. In terms of the spatial distribution 
of the 511 keV line component, the early mapping results point to an 
intense bulge emission, better explained by an extended 
distribution than by a point source. The disk component has been either 
absent, or weakly detected in the initial results. The spectrum has 
revealed that the line centroid is centered at 511 keV with a line profile 
that can be separated into two components. These findings have motivated 
speculation as to the source(s) of the bulk of galactic positrons, 
both through further study of previously suggested sources, and through 
the development of new mechanisms. Collectively, the findings from 
SPI and the speculation that it has generated makes this a very interesting 
time for positron annihilation astrophysics. 

The impression exists that the SPI findings are in disagreement with 
the spatial distributions reported by the previous missions, most 
notably by the CGRO/OSSE instrument. In particular, the total flux 
reported in many SPI-based publications is lower than the total fluxes 
reported by Purcell et al. 1997 and Milne et al. 2002, based upon 
OSSE/SMM/TGRS observations. Many SPI-based publications report no 
statistical requirement for the existence of 
a disk component to explain their observations. 
It is critical for the advancement of positron annihilation astrophysics 
to determine whether this impression of disagreement is correct, or 
rather a false impression generated by differing points-of-emphasis 
between the different teams presenting results from their instruments. 


In this paper, we present a few simple comparisons between SPI results 
and those reported by OSSE, SMM and TGRS. In all cases, we have tried to
compare similar models to the extent possible from the literature. We 
emphasize that this is merely the first step of a collaborative effort 
to jointly study the observations made by the OSSE and SPI instruments. 
In some situations, it works out that the improvement from one generation 
of instrument to the next generation is large enough that the newer 
instrument completely redefines the level of understanding of a 
particular emission. This is not the case for positron annihilation 
radiation. It works out that the best description of the distribution 
and total flux of positron annihilation radiation will utilize the 
observations performed by both instruments, with the potential that 
the joint analysis will lead to better direction as to how to best 
observe annihilation radiation with SPI. 

\begin{figure}[tb]
\begin{center}
\includegraphics[width=.8\columnwidth]{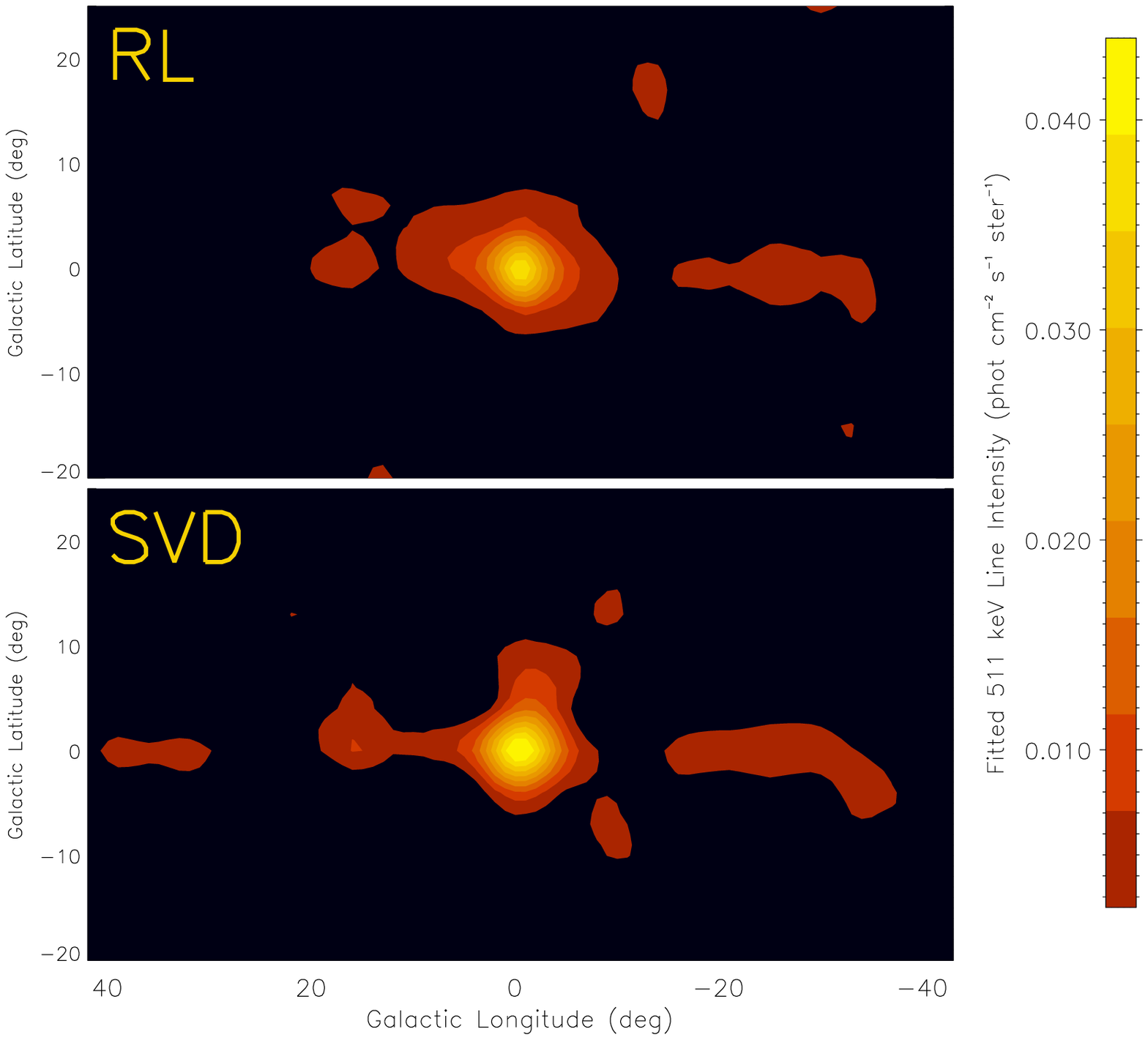}
\includegraphics[width=.8\columnwidth]{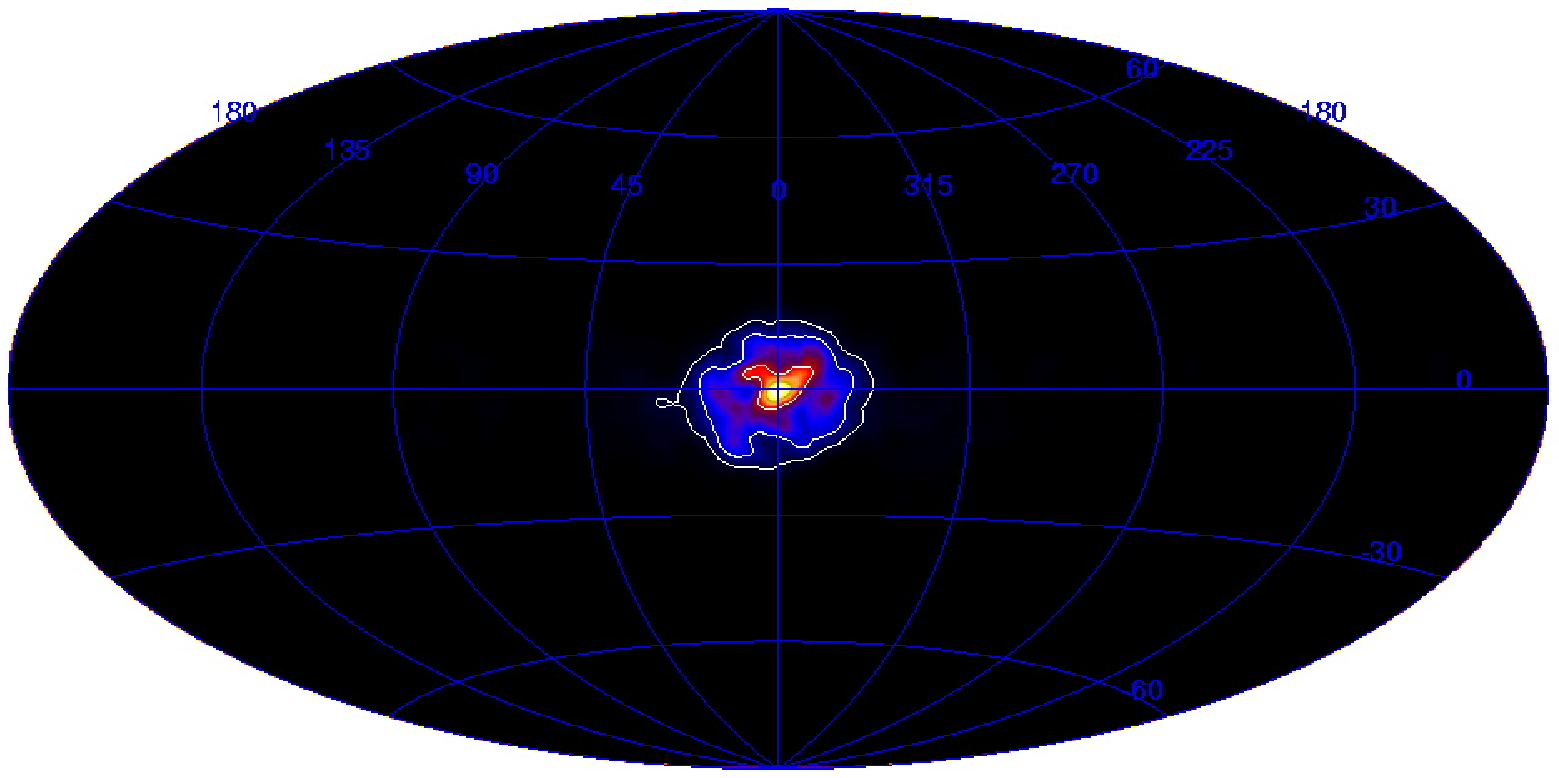}
\caption{Maps of 511 keV line emission. The upper panels show 
maps made from OSSE/SMM/TGRS observations with an adaptation 
of the Richardson-Lucy Chi-square minimization algorithm 
(top) and with SVD response matrix inversion (middle). The 
lower panel shows a map made from SPI observations (Kn\"{o}dlseder 
et al. 2005). Lines on SPI map follow contours of 
10$^{-4}$,10$^{-3}$,10$^{-2}$ phot cm$^{-2}$~s$^{-1}$sr$^{-1}$.}
\label{ost511}
\end{center}
\end{figure}

\section{SMM, TGRS and OSSE Observations}

The SMM instrument observed positron annihilation radiation during the 
1980s. The SMM instrument had relatively poor energy resolution 
(41 keV @ 511 keV), but a wide FoV ($\sim$130 deg), which permitted 
long-term monitoring of the galactic center region during its nine 
years of operation. Principal among the SMM findings is the lack of 
significant variation in the 511 keV line flux from the direction of 
the galactic center, and an estimate of the total flux of 511 keV line 
emission, between (1.6 - 3.0)~x~10$^{-3}$~phot~cm$^{-2}$s$^{-1}$, depending 
upon the spatial distribution of the emission (Share et al. 1988). 
SMM observations afford very little 
insight into the distribution of annihilation radiation, but they suggest  
a total flux level once a distribution is assumed. 

The TGRS instrument observed positron annihilation radiation during the 
early-to-mid 1990s. With a germanium spectrometer, TGRS was capable of 
studying the 511 keV line centroid, thickness and profile, as well as 
being able to estimate the positronium continuum flux. Also a wide 
FoV instrument, TGRS reinforced the lack of variability of the annihilation 
flux, in this case during the mid-1990s. 

The OSSE instrument observed positron annihilation radiation during the 
1990s, utilizing a tungsten collimator over CsI(Th) detectors. 
OSSE had a much narrower FoV than SMM and TGRS 
(3.8$^{\circ}$~x~11.4$^{\circ}$), and thus had a 
far superior ability to determine the spatial distribution of annihilation 
radiation. It also accomplished background supression through alternating 
between source and background pointings every two minutes. This approach 
made the observations fundamentally differential rather than absolute, 
but it lowered the dependence upon understanding the instrumental background 
to derive accurate results. OSSE had a relatively poor energy resolution, 
(45~keV~@~511~keV), and thus contributed little to the understanding 
of the 511 keV line centroid, thickness and profile.\footnote{OSSE would 
have been capable of measuring significantly red-shifted or broadened 
annihilation radiation, however, analysis of OSSE observations has not 
revealed evidence of this emission.} Both the 511 keV line 
and the positronium continuum fluxes have been mapped, the first time maps 
have been created of annihilation radiation.

The OSSE instrument was the best capable of studying the distribution 
of annihilation radiation, and OSSE continued to observe annihilation 
radiation after SMM and TGRS had completed the annihilation radiation 
portions of their missions. However, it has been determined that for a 
large subset of possible spatial distributions, the data-sets for the 
OSSE, SMM and TGRS instruments could be acceptably fitted. 
Mapping and model-fitting results for 511 keV line emission were often 
presented for all three data-sets simultaneously (the positronium continuum 
emission maps were generated from OSSE data alone). One principal finding 
from OSSE studies is that the 511 keV line and positronium continuum 
emission are similarly distributed, with both emissions featuring an 
intense bulge component and a fainter disk component. 

The SPI instrument features spatial resolution  
(2.5$^{\circ}$~x~2.5$^{\circ}$) and spectral resolution ($\sim$2 keV) 
far superior to that achieved by the OSSE instrument. Employing a 
coded mask above a Ge spectrometer, SPI accomplishes imaging and 
suppression of background in a different manner than OSSE. These 
differences lead to a complementary situation where each instrument's 
findings are largely independent of the systematics of the other 
instrument. I this work, we cite the SPI study of Kn\"{o}dlseder 
et al. 2005 for the spatial distribution of 511 keV line emission; 
Lonjou et al. 2004 and Jean et al. 2006 for line diagnostics 
of the 511 keV line emission; 
and Weidenspointner et al. 2006 (in preparation) for positronium 
continuum emission. 

\section{Comparisons and Discussion}

OSSE/SMM/TGRS maps of 511 keV line emission are shown in the upper 
panel of Figure \ref{ost511}. The intense bulge emission is evident, 
and a fainter disk emission is also present. The lower panel shows 
a SPI map of 511 keV line emission. The intense bulge emission is 
again evident, but the disk emission is absent. The disk is not 
apparent in SPI maps because it is either too faint and/or too 
longitudinally thick to be detectable with the current set of 
SPI observations. The question is then whether the disk emission 
apparent in OSSE/SMM/TGRS maps is in conflict with the 
lack of disk emission in SPI maps. This question is best addressed 
through comparisons of model-fitting to each data-set.

\begin{figure}[tb]
\begin{center}
\includegraphics[width=.9\columnwidth]{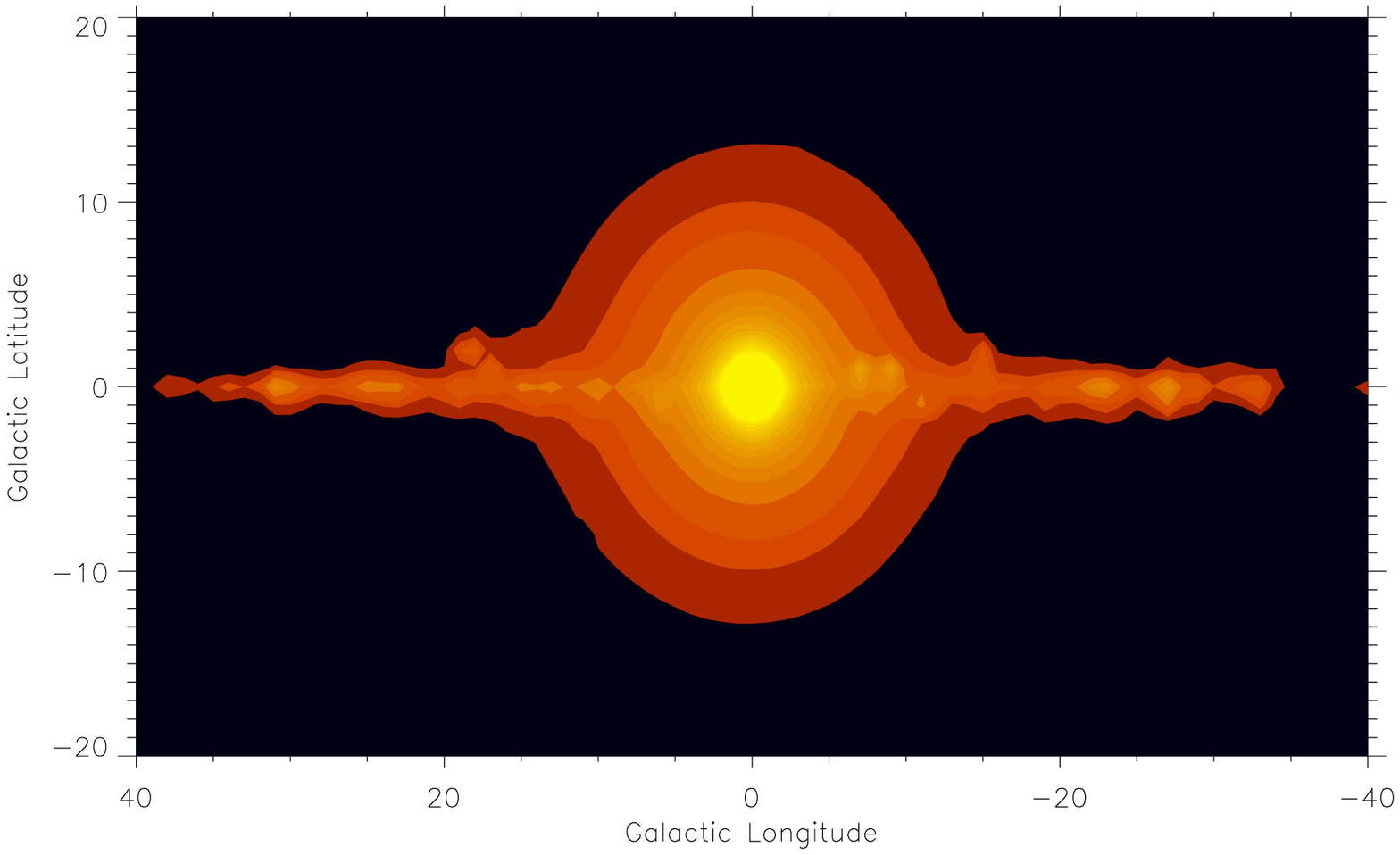}
\includegraphics[width=.9\columnwidth]{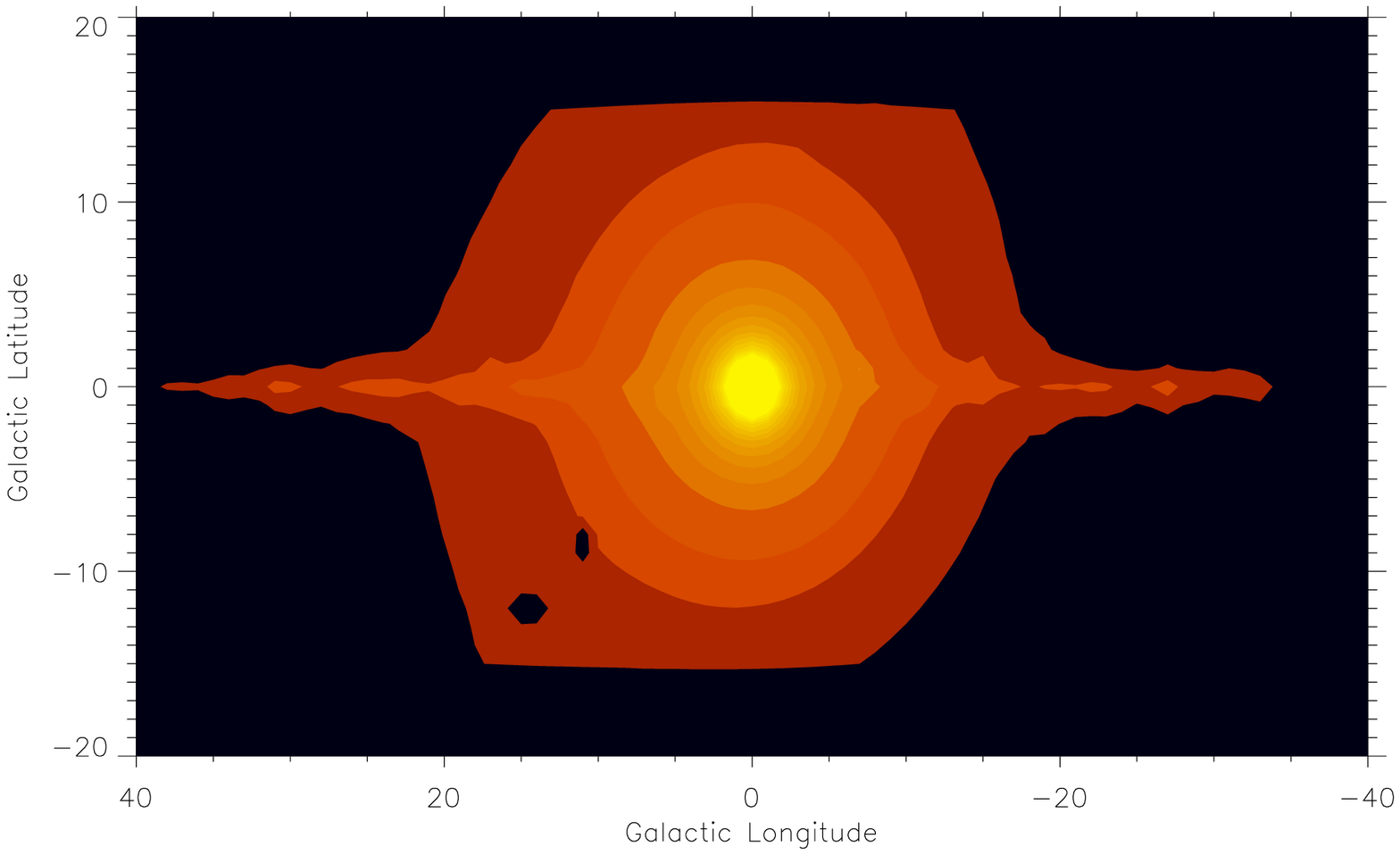}
\includegraphics[width=.9\columnwidth]{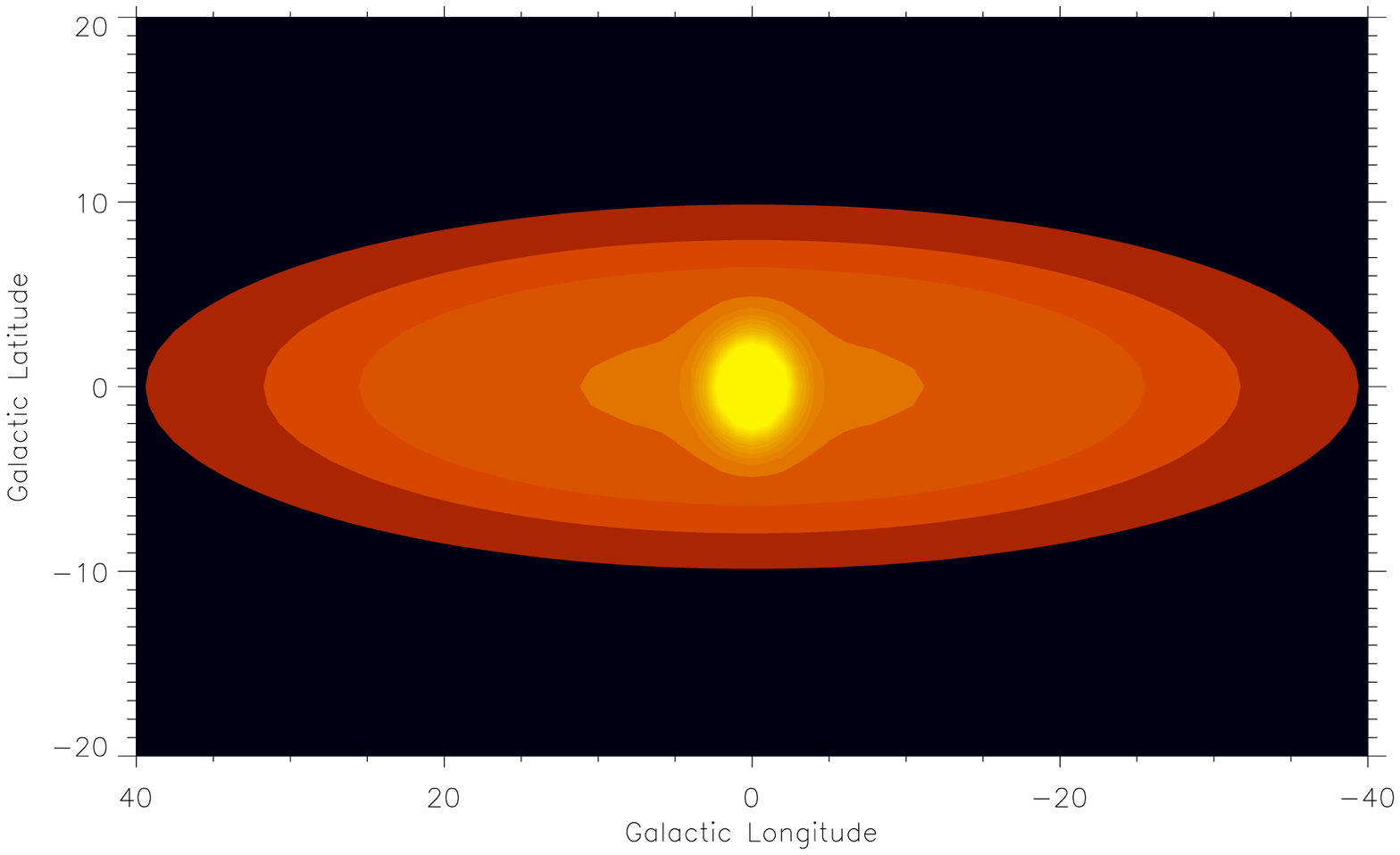}
\caption{Three bulge-disk pairs that are equally-acceptable explanations 
for OSSE/SMM/TGRS observations of positron annihilation radiation. The 
upper two panels feature an R$^{1/4}$ spheroid, the lower panel features 
a Gaussian spheroid. The latitudinal thickness of the disk increases from 
the upper panel to the lower panel. Contours are on the same scale as 
Figure 1.} 
\label{bdpair}
\end{center}
\end{figure}

As described during AwR4 (Milne, Kurfess, Kinzer \& Leising 2002) 
the OSSE data (both 511 keV line and positronium 
continuum) are equally well-fitted by three families of bulge-disk pairs; 
a bulge-dominated solution, an equal-contribution solution; and a 
disk-dominated solution. Examples of each of these three pairings 
are shown in Figure \ref{bdpair}. Although these exact model pairings 
were not compared with the SPI data in the Kn\"{o}dlseder study, 
rough matches can be found. The ``H-Prime" spheroid and the R$^{1/4}$ 
spheroid possess similar shapes. The ``Shells" spheroid and the 
2D-Gaussian spheroid are quite similar. The ``D0-young" disk in 
a thin disk, while the ``D1-old" disk is a thicker disk.  
Shown in Table 1 are bulge, disk and total fluxes of bulge-disk pairs fit to 
OSSE, OSSE-SMM, OSSE-TGRS data and equivalent pairs fit to SPI data. It is 
clear from the table that when similar bulge-disk shape are used, 
OSSE/SMM/TGRS and SPI arrive at similar fluxes.  As the INTEGRAL mission 
continues, the significance of the disk component should increase. In 
addition, SPI observations should eventually be capable of eliminating 
two of the ``OSSE allowed" bulge-disk families. 

\begin{table}
\begin{center}
\begin{tabular}{lccc}
Model & Bulge & Disk & Total \\
\hline
R1/4 - Thin Disk & & & \\
OSSE & 1.7 & 0.3 & 2.0 \\
OSSE-SMM & 2.0 & 0.4 & 2.4 \\
OSSE-TGRS & 1.6 & 0.3 & 1.9 \\
OSSE-SMM-TGRS & 2.5 & 0.3 & 2.8 \\
{\it SPI: H' + D0} & 1.6 & 0.4 & 2.1 \\
\hline
R1/4 - Thick Disk & & & \\
OSSE & 1.9 & 0.9 & 2.8 \\
OSSE-SMM & 1.8 & 1.0 & 2.8 \\
OSSE-TGRS & 2.0 & 0.6 & 2.6 \\
OSSE-SMM-TGRS & 1.8 & 0.7 & 2.6 \\
{\it SPI: H' + D1} & 1.5 & 0.9 & 2.4 \\
\hline
Gaussian - Thick Disk & & & \\
OSSE & 0.5 & 0.5 & 1.0 \\
OSSE-SMM & 0.3 & 2.2 & 2.5 \\
OSSE-TGRS & 0.8 & 0.7 & 1.6 \\
OSSE-SMM-TGRS & 0.3 & 2.2 & 2.5 \\
{\it SPI: Shells  + D1} & 1.0 & 1.0 & 2.0 \\
\hline
\end{tabular}
\caption{Fluxes of Bulge-Disk Pairs (10$^{-3}$~phot~cm$^{-2}$~s$^{-1}$).}
\end{center}
\end{table}

OSSE, TGRS and SPI have all reported positronium fractions (i.e. the 
fraction of positron-electron annihilations that proceed after forming 
Positronium). Using TGRS data, Harris et al. 1998 reported 
f(Ps)=0.94$\pm$0.04. Using OSSE data, Kinzer et al. 2001 reported 
f(Ps)=0.97$\pm$0.04. Using SPI data, Weidenspointner reports 
f(Ps)=0.93$\pm$0.09, which is in agreement with the other two values 
(personal communication). 

TGRS and SPI have studied the line profile of the 511 keV line. 
Using TGRS data, Harris et al. 1998 reported the centroid at 
510.98~$\pm$~0.10~$\pm$~0.14 keV, a narrow-component line width of 
1.81~$\pm$~0.54~$\pm$~0.14 keV, and a broad line component with 
11\%~$\pm$~9\% of the total flux. Using SPI data, Lonjou et al. 2004 
report the line to be centered at 511.02~+0.08~-0.09 keV. Also using 
SPI data, Jean et al. 2006 report a  
narrow-component line width of 1.3~$\pm$~0.4 keV FWHM, and a broad-line 
component with 33\%~$\pm$~11\% of the total flux. While the TGRS and 
SPI results are not identical, they do show a high level of agreement.  

Viewed collectively, the SPI results re-affirm the essential findings from 
the SMM, TGRS and OSSE observations. As the INTEGRAL mission should be 
operational for years to come, the potential is great for SPI to 
further advance the characterization of galactic positron annihilation, 
and thus positron astrophysics in general. When comparing values, 
such as bulge, disk and total fluxes, B/D ratios, and line widths, it 
is important to be certain that the comparison is between similar 
models or assumptions. Apparent discrepancies might be nothing more than 
mismatches between the models being compared. A joint effort to properly 
compare OSSE/SMM/TGRS results with SPI results is underway, and promises 
to produce the clearest picture yet of the distribution of positron 
annihilation radiation in the Galaxy.


\begin{thebibliography}{}
\bibitem{harr98}Harris,M.~J. et al. 1998, {\it ApJ}, {\bf 501}, L55. 
\bibitem{jean06}Jean, P. et al. 2006, {\it A\&A} {\bf 445}, 579.
\bibitem{kinz01}Kinzer, R.L. et al. 2001, {\it Ap. J.}, {\bf 559}, 282.
\bibitem{knod05}Kn\"{o}dlseder et al. 2005, {\it A\&A} {\bf 441}, 513. 
\bibitem{lonj04}Lonjou, V. et al. 2004, in ``The INTEGRAL Universe", 
Proc. of the 5th INTEGRAL Workshop held 16-20 February 2004, Munich, 
Germany, ESA SP-552, 129. 
\bibitem{miln02}Milne, P.A., Kurfess, J.D., Kinzer, R.L., Leising, M.D. 2002,
{\it NewAR} {\bf 46}, 553.
\bibitem{purc97}Purcell, W.~R., et~al., 1997, ApJ, 491, 725. 
\bibitem{shar88}Share, G.H. et al. 1988, {\it ApJ} {\bf 326}, 717. 
\end{thebibliography}
\end{document}